\documentclass[11pt]{article}
\usepackage{amsmath,latexsym}
\usepackage{graphicx}
\begin{document}

\title{\textbf{\textsf{Model of interacting dark energy in five-dimensional brane cosmology}}}
\author{ \large \textbf{\textsf{Mubasher Jamil}}\footnote{Email: mjamil@camp.edu.pk}
\\ \\
\small\textit{Center for Advanced Mathematics and Physics}\\
\small\textit{National University of Sciences and Technology}\\
\small\textit{E\&ME campus, Peshawar road, Rawalpindi - 46000, Pakistan}\\
}\maketitle

\begin{abstract}
We here investigate the model of interacting dark energy in the
context of five dimensional brane cosmology. The effective equations
of state of dark energy are evaluated for various choices of the
variable time dependent cosmological constant. We have found that
the interacting dark energy obeys the phantom divide/crossing
scenario in this generalized model. It is also shown that
interacting dark energy in this generalized model also resolves the
cosmic coincidence problem.

\end{abstract}


\section{Introduction}

Recent astrophysical observations give bunch of convincing evidence
of our universe undergoing accelerated expansion
\cite{perlmutter1,perlmutter3,wang,riess1,riess2,spergel,tonry,gong}.
A perturbing feature of this phenomenon is that it was preceded by a
decelerated expansion, so we need to ask what caused this sudden
transition \cite{caldwell1}. Observations also show that this shift
is rather marginally recent (less then one Gyr), hence it poses why
it happened so recently. If we assume that the source which is
driving this expansion is some mysterious `dark energy' then one
needs to ask what is the composition of this exotic matter, also why
it has become dominating all of a sudden at present time (see
\cite{ratra,sahni,weinberg,copeland,padmanabhan} for reviews on dark
energy). In other words, why it was negligibly small in earlier
deceleration phase. Investigations of WMAP show that energy
densities of dark energy and matter are almost comparable at present
time. This leads to a problem named as the `cosmic coincidence
problem' (CCP). The problem aggravated when it was shown that
radiation energy density was also equivalent to that of dark energy,
leading to a `cosmic triple coincidence problem' \cite{hamed},
however it is recently addressed in the context of triple
interacting fluids \cite{jamil2}. Further if the universe is dark
energy dominated, will it expand forever or may decelerate at some
instant as well.

The dark energy is generally represented by a phenomenologically
motivated equation of state (EoS) $p_{de}=\omega_{de}\rho_{de}$,
where $p_{de}$ and $\rho_{de}$ are the pressure and the energy
density of the dark energy, interlinked by a dimensionless parameter
$\omega_{de}$ \cite{sean}. In order to produce the accelerated
expansion, it requires $\omega_{de}<-1/3$. In recent years, several
theoretical models have been proposed to understand the nature and
dynamics of dark energy, however almost all these models either
require fine tuning of their model parameters or yield quantum or
gravitational instabilities that are needed to be removed. Most
prominent dark energy proposals are based on cosmological constant
$\Lambda$ \cite{paul,ishak}, quintessence \cite{huang,bharat},
k-essence \cite{das,picon}, phantom energy \cite{caldwell2}, quintom
model \cite{cai1,cai2}, geometric dark energy \cite{eric},
holographic dark energy \cite{setare1} and tachyons
\cite{padmanabhan1,diaz}, to name a few. It has been pointed out
that quantum effects can yield a super-accelerated phase of cosmic
expansion (without any need of introducing ghosts, phantoms or
tachyons) and that these quantum effects yield stable solutions
\cite{Onemli1,Onemli2,Onemli3}. The precise determination of
$\omega_{de}$ is a more challenging and interesting problem in
itself. Recent observational data gives the estimate
$-1.67<\omega_{de}<-1.05$ at $95\%$ confidence level
\cite{macorra1}. It also supports the notion of an evolving
$\omega_{de}$, hence it requires the parametric form
$\omega_{de}(z)$, where $z$ is the redshift parameter
\cite{polarski,barboza,weller,linder,efstathiou,upadhye}. For larger
redsifts (quintessence dominated), $\omega_{de}>-1$ while at some
instant $\omega_{de}=-1$ ($\Lambda$ dominated era) and later it is
$\omega_{de}<-1$ (phantom regime). Therefore the CCP is rephrased as
`why now $\omega_{de}=-1?$'

In the last few years, the CCP is addressed by invoking a
non-minimal interaction between dark energy and dark matter (or
simply `matter' for convenience), the so-called interacting dark
energy model
\cite{wen,marek,barrow,macorra,neto,setare,nojiri,winfried,cimento,tame,mota,banerjee,cruz}.
The model involves a coupling parameter and an energy exchange term
to govern the interaction. The energy exchange term is adjusted so
as to satisfy the global conservation law for the interacting
system. Moreover the interaction is dynamic i.e. the energy is
exchanged between the interacting components with equal degree of
freedom. It is exactly this feature that helps in maintaining the
equilibrium of densities of the interacting components i.e. the
ratio of energy densities roams around to unity. It is recently
shown that the coupling parameter cannot be negative in order to
avoid possible violation of the cherished second law of
thermodynamics, however, small positive values are permissible to
account the decay of dark energy into matter
\cite{feng1,li,jamil22}. The notion of decay of dark energy into
matter particles is well motivated from the theoretical arguments
and henceforth predicts a matter dominated universe. Hence a
universe governed by the interacting components can undergo
deceleration phase preceded by an acceleration phase. Further, the
interaction also saves the universe from undergoing an imminent `big
rip' (tearing apart of spacetime structure leading to a future
spacelike singularity). This model also favors a bouncing universe
(free from cosmological singularities like big bang, big rip, big
crunch etc) since the model forces the components to interact,
thereby avoiding cosmological over-densities.

We here extend our earlier work on interacting dark energy
\cite{jamil,jamil1} in the context of five dimensional brane-world
model \cite{gron}. Last few decades have seen a considerable
advancement on the theories of extra dimensions and have revealed
deep insights about the structure of spacetime, elementary particles
and forces of nature \cite{polch}. It has been suggested that the
mysterious dark energy is also a manifestation of extra spatial
dimensions \cite{ito}. We investigate the behavior of dark energy in
the brane world model and determine various EoSs for the dark energy
for different choices of the time dependent cosmological constant.
All the effective EoS describe the phantom crossing scenario under
certain conditions. It is also discussed that this generalized
interacting dark energy model fairly resolves the CCP.

\section{Interacting dark energy model}

We start by assuming the background to be spatially homogeneous and
isotropic Friedmann-Robertson-Walker (FRW) spacetime, in the context
of brane-world gravitation model, which is given by
\cite{kamdi,gron}
\begin{equation}
ds^2=dt^2-a^2(t)\left[
\frac{dr^2}{1-kr^2}+r^2(d\theta^2+\sin^2\theta d\phi^2)
\right]-\zeta^2(t)d\psi^2.
\end{equation}
Here $k$ is a curvature parameter which refers to a spatially
spherical ($k=+1$), Minkowskian ($k=0$) or hyperbolic ($k=-1$)
spacetime. Note that these cosmological models are also correspond
to closed, flat or open respectively. Also $a(t)$ is the
dimensionless scale factor while $\zeta$ is an arbitrary function of
time $t$ and we choose it $\zeta(t)=a^n$, whereas $n$ is a constant.
Moreover $\psi$ is the fourth spatial dimension. The spacetime is
further assumed to contain two fluids namely matter and dark energy.
The corresponding energy densities are $\rho_m$ and $\rho_{de}$
while the respective pressures are $p_m=0$ (pressureless dust) and
$p_{de}=\omega_{de}\rho_{de}\neq0$. The combined matter energy
distribution is given by a perfect fluid stress energy tensor
\begin{equation}
T_{\mu\nu}=(\rho_m+\rho_{de}+p_{de})u_\mu u_\nu-p_{de} g_{\mu\nu}.
\end{equation}
Here $u_\mu$ is the five-velocity vector which satisfies $u_\mu
u^\nu=1$ with $\mu,\nu=0,1,2,3,4$ . The first FRW equation is
\begin{equation}
(n+1)H^2+\frac{k}{a^2}
=\frac{1}{3M^2_p}(\rho_{de}+\rho_m)+\frac{\Lambda(t)}{3}.
\end{equation}
Here $M_p^2=(8\pi G)^{-1}$ is the reduced Planck mass. The
corresponding energy conservation equation is
\begin{equation}
\dot{\rho}_{de}+\dot{\rho}_m+(3+n)[\rho_m+\rho_{de}(1+\omega_{de})]H=-M_p^2\dot{\Lambda},
\end{equation}
which can be decomposed into two non-conserving equations for both
matter and dark energy as
\begin{eqnarray}
\dot{\rho}_{de}+(3+n)(1+\omega_{de})\rho_{de}
H&=&-M_p^2\frac{\dot{\Lambda}}{2}-Q,\\
\dot{\rho}_m+(3+n)\rho_m H&=&-M_p^2\frac{\dot{\Lambda}}{2}+Q.
\end{eqnarray}
Note that addition of the above two equations leads to the energy
conservation (4). Above $Q$ is the energy exchange term for the
interaction. We here choose $Q=3Hb(\rho_m+\rho_{de})$
\cite{guo,campo22}, where $b$ is the coupling parameter (or transfer
strength). Due to unknown nature of both dark energy and dark
matter, the interaction term can not be derived from the first
principles. It is worthy to note that if $Q<0$ than it will yield
the energy density of dark energy to be negative at sufficiently
early times, consequently the second law of thermodynamics can be
violated \cite{lima} hence $Q$ must be positive and small. Because
of the underlying interaction, the beginning of the accelerated
expansion is shifted to higher redshifts.

Further the density parameters are
\begin{eqnarray}
\Omega_m&=&\frac{\rho_m}{\rho_{cr}}=\frac{\rho_m}{3H^2M^2_p},\\
\Omega_{de}&=&\frac{\rho_{de}}{\rho_{cr}}=\frac{\rho_{de}}{3H^2M^2_p},\\
\Omega_k&=&\frac{k}{a^2H^2}.
\end{eqnarray}
Using the above parameters in Eq. (3), we obtain
\begin{equation}
\Omega_m+\Omega_{de}=1+n+\Omega_k-\frac{\Lambda(t)}{3H^2}.
\end{equation}
We here define the dimensionless ratio of densities
\begin{equation}
r_x\equiv\frac{\rho_m}{\rho_{de}}=\frac{\Omega_m}{\Omega_{de}}.
\end{equation}
To check how this density ratio evolves with time, we differentiate
it w.r.t $t$ to get
\begin{equation}
\dot
r_x=\frac{dr_x}{dt}=\frac{\rho_m}{\rho_{de}}\left[\frac{\dot\rho_m}{\rho_m}-\frac{\dot\rho_{de}}{\rho_{de}}
\right]\equiv f(r_x).
\end{equation}
Using Eqs. (5) and (6) in (12), we obtain
\begin{equation}
\dot r_x=r_x(3+n)H\left[\omega_{de}
+\frac{\Gamma}{(3+n)H}\frac{1+r_x}{r_x}+\frac{\dot\Lambda(r_x-1)}{6H^3\Omega_m(3+n)}
\right],
\end{equation}
where
\begin{equation}
\Gamma=3Hb(1+r_x),
\end{equation}
is the decay rate, related to $Q=\Gamma\rho_{de}$. Thus as $r_x$
approaches 1, the last term on right hand side in equation (13)
becomes negligible. It is also termed as the `soft coincidence'
since $|\dot r_x/r_x|\leq H$ \cite{campo}. Further if $n=0$ then Eq.
(13) reduces to the one discussed in \cite{setare}. Using Eq. (14)
in (13), we obtain
\begin{equation}
\dot r_x=r_x(3+n)H\left[\omega_{de}
+\frac{3b(1+r_x)^2}{r_x(3+n)}+\frac{\dot\Lambda(r_x-1)}{6H^3\Omega_m(3+n)}
\right].
\end{equation}
The critical points (or stationary solutions) are obtained by
solving $\dot r_x=0$ to get
\begin{equation}
r_x^2\left[ \dot\Lambda+18bH^3\Omega_m \right]+r_x\left[
6H^3\Omega_m\{6b+\omega_{de}(3+n)\}-\dot\Lambda
\right]+18bH^3\Omega_m=0.
\end{equation}
The above equation yields two roots as
\begin{eqnarray}
r_{x\pm}&=&\frac{1}{2(\dot\Lambda+18bH^3\Omega_m)}[\dot\Lambda-6H^3\Omega_m\{6b+\omega_{de}(3+n)\}\nonumber\\&\;&\pm
\sqrt{-72H^3\Omega_mb(\dot\Lambda+18bH^3\Omega_m)+\{\dot\Lambda-6H^3[6b+(3+n)\omega_{de}]\Omega_m\}^2}].
\end{eqnarray}
It is recently shown in \cite{tame} that any model of interacting
dark energy can resolve the cosmic coincidence problem if the
function $f(r_x)$ satisfies
\begin{equation}
\frac{df}{dr}(r={r_x}_i)<0,
\end{equation}
where ${r_x}_i$ for $i=1,2,...$ are the roots of $f(r_x)=0$. It
needs to be stressed that not all roots will satisfy (18) but those
which do satisfy it, are termed `stable equilibrium points'. In our
model, the condition (18) together with (17) yields
\begin{equation}
f'(r_{
x\pm})=\pm\frac{\sqrt{-72H^3\Omega_mb(\dot\Lambda+18bH^3\Omega_m)+[\dot\Lambda-6H^3\{6b+(3+n)\omega_{de}\}\Omega_m]^2}}{6H^2\Omega_m}.
\end{equation}
Hence it is clear that $f'(r=r_{x-})<0$ and $r_{x-}$ is the only
stable equilibrium point of our model. Thus interacting dark energy
model in brane-world gravitation theory fairly alleviates the CCP.
Our next task is now to determine the effective EoS for dark energy.

The parameter $r_x$ in (11) is related to the density parameters (7)
- (9) as
\begin{equation}
r_x=\frac{1}{\Omega_{de}}\left[1+n+\Omega_k-\Omega_{de}-\frac{\Lambda(t)}{3H^2}
\right].
\end{equation}
We further define the effective equations of state for dark energy
and matter as \cite{setare}
\begin{equation}
\omega_{de}^{eff}=\omega_{de}+\frac{\Gamma}{3H}, \ \
\omega_m^{eff}=-\frac{1}{r_x}\frac{\Gamma}{3H},
\end{equation}
From Eq. (5) we have
\begin{equation}
\omega_{de}=-1-\frac{1}{(3+n)H\rho_{de}}\left[
Q+\dot{\rho}_{de}+M_p^2\frac{\dot{\Lambda}}{2} \right].
\end{equation}
Using Eq. (22) in (21), we get
\begin{equation}
\omega_{de}^{eff}=-1-\frac{1}{(3+n)H\rho_{de}}\left[
\dot{\rho}_{de}+M_p^2\frac{\dot{\Lambda}}{2}
\right]+\frac{n\Gamma}{3(n+3)H},
\end{equation}
or we can write
\begin{equation}
\omega_{de}^{eff}=-1-\frac{\dot\rho_{de}}{(3+n)H\rho_{de}}-\frac{\dot\Lambda}{6(3+n)H^3\Omega_{de}}+\frac{nb(1+r_x)}{n+3}.
\end{equation}
We represent the dark energy by the modified Chaplygin gas (MCG)
equation of state
\begin{equation}
p_{de}=A\rho_{de}-\frac{B}{\rho_{de}^\alpha},
\end{equation}
where $A$, $B$ and $\alpha$ are constant parameters. The MCG best
fits with the $3-$year WMAP and the SDSS data with the choice of
parameters $A=-0.085$ and $\alpha=1.724$ \cite{lu} which are
improved constraints than the previous ones $-0.35<A<0.025$
\cite{jun}. Recently it is shown that the dynamical attractor for
the MCG exists at $\omega_{de}=-1$, hence MCG crosses this value
from either side $\omega_{de}>-1$ or $\omega_{de}<-1$, independent
to the choice of model parameters \cite{jing}. A generalization of
MCG is suggested in \cite{debnath} by considering $B\equiv
B(a)=B_oa^\sigma$, where $\sigma$ and $B_o$ are constants. The MCG
is the generalization of generalized Chaplygin gas
$p_{de}=-B/\rho_{de}^\alpha$ \cite{sen,carturan} with the addition
of a barotropic term. This special form also appears to be
consistent with the WMAP $5-$year data and henceforth the support
the unified model with dark energy and matter based on generalized
Chaplygin gas \cite{barriero,makler}. In the cosmological context,
the Chaplygin gas was first suggested as an alternative to
quintessence and demonstrated an increasing $\Lambda$ behavior for
the evolution of the universe \cite{kamenshchik}. Recent supernovae
data also favors the two-fluid cosmological model with Chaplygin gas
and matter \cite{grigoris}. The density evolution of MCG is given by
\begin{equation}
\rho_{de}=(X+C_1a^Y)^{\frac{1}{1+\alpha}},
\end{equation}
where $X\equiv B/(1+A)$, $Y\equiv3(1+\alpha)(1+A)$ and $C_1$ is the
constant of integration. The time derivative of $\rho_{de}$ is given
by
\begin{equation}
\dot{\rho}_{de}=-3(1+A)C_1H(X+C_1a^Y)^{\frac{-\alpha}{1+\alpha}}a^Y.
\end{equation}
Using Eqs. (24) to (27), we obtain the effective EoS for dark energy
as
\begin{equation}
\omega_{de}^{eff}=-1+\frac{3(1+A)}{3+n}\left(
1-\frac{X}{\rho_{de}^{1+\alpha}}
\right)-\frac{\dot\Lambda}{6(3+n)H^3\Omega_{de}}+\frac{nb(1+r_x)}{n+3}.
\end{equation}
Models with variable $\Lambda(t)$ are physically more appealing and
theoretically rich in predictions as compared to constant $\Lambda$.
If $\dot{\Lambda}<0$, it gives a decreasing behavior of $\Lambda$
with time. Physically it may explain inflationary expansion at
earlier times while an accelerated expansion in current time.
Similarly, $\dot{\Lambda}>0$ represents an increasing $\Lambda$ with
time. It can be best interpreted in a model of bouncing cosmology
where a universe is free from any potential cosmological
singularities like the big bang one and bounces back near the
imminent singularity. In such a scenario, a smaller $\Lambda$
corresponds to a deceleration phase followed by a smoothly evolving
larger $\Lambda$ which results in a de Sitter like expansion. If we
assume that this later expansion is driven by an exotic phantom
energy ($\omega_{de}<-1$), then the later one decays into matter
particles creating a matter dominated universe again \cite{curbelo}.
Thus in a bouncing universe, an otherwise big rip singularity is
replaced by the matter creation scenario. Hence if $\omega_{de}<-1$,
it eventually leads to two interesting results: first, the existence
of a bouncing universe and second, the decay of dark energy into
matter or the model of interacting dark energy \cite{anton,neto}. It
is recently suggested using inhomogeneous EoS for dark energy that
the dark energy dilution becomes faster in de Sitter expansion which
involves strong interaction between dark energy and matter
\cite{writ}. We shall now proceed to determine $\omega_{de}^{eff}$
by assuming dependencies of $\Lambda$ on various cosmological
parameters and determine conditions under which it will become
super-negative.

From Eq. (28) we see that $\omega_{de}^{eff}<-1$ if
$\dot{\Lambda}>0$ and $n>-3$. The scenario of dark energy dilution
into matter arises for $b>0$ or $b\rightarrow1$ if we restrict
$0\leq b\leq1$ while a $b=0$ corresponds to a non-interacting dark
energy model. It implies that for some specific values like
$n=-1,-2$ the last term in Eq. (28) will also be negative. The case
for $b<0$ refers to matter decay into dark energy, which is not
relevant here. Let us choose $\Lambda(t)=C_2t^{\beta}$ than Eq. (28)
yields
\begin{equation}
\omega_{de}^{eff}=-1+\frac{3(1+A)}{3+n}\left(
1-\frac{X}{\rho_{de}^{1+\alpha}} \right)-\frac{C_2\beta
t^{\beta-1}}{6(3+n)H^3\Omega_{de}}+\frac{nb(1+r_x)}{n+3}.
\end{equation}
Next we choose $\Lambda(t)=C_3e^{\gamma t}$ which gives
\begin{equation}
\omega_{de}^{eff}=-1+\frac{3(1+A)}{3+n}\left(
1-\frac{X}{\rho_{de}^{1+\alpha}} \right)-\frac{C_3\gamma e^{\gamma
t}}{6(3+n)H^3\Omega_{de}}+\frac{nb(1+r_x)}{n+3}.
\end{equation}
Thus in the above two Eqs. (29) and (30), $\dot{\Lambda}>0$
translates into $\beta>1$ and $\gamma>0$. Next we take
$\Lambda(t)=C_4a^{\delta}$ which enables us to write
\begin{equation}
\omega_{de}^{eff}=-1+\frac{3(1+A)}{3+n}\left(
1-\frac{X}{\rho_{de}^{1+\alpha}} \right)-\frac{C_4\delta
a^{\delta}}{6(3+n)H^2\Omega_{de}}+\frac{nb(1+r_x)}{n+3}.
\end{equation}
which can alternatively be written as
\begin{equation}
\omega_{de}^{eff}=-1+\frac{3(1+A)}{3+n}\left(
1-\frac{X}{\rho_{de}^{1+\alpha}} \right)-\frac{C_4\delta
\left[\frac{1}{C_1}\left(\rho_{de}^{1+\alpha}-X
\right)\right]^{\delta/Y}}{6(3+n)H^2\Omega_{de}}+\frac{nb(1+r_x)}{n+3}.
\end{equation}
Here we require $\delta>0$ and $\delta>Y$ to get a super-negative
EoS. If we take $\Lambda(t)=C_5H^\upsilon$ then we have
\begin{equation}
\omega_{de}^{eff}=-1+\frac{3(1+A)}{3+n}\left(
1-\frac{X}{\rho_{de}^{1+\alpha}} \right)-\frac{C_5\upsilon
H^{\upsilon-4}\dot H}{6(3+n)\Omega_{de}}+\frac{nb(1+r_x)}{n+3}.
\end{equation}
The phantom crossing scenario is more prominently observed from
$\dot H>0$ ($\omega_{de}<-1$), $\dot H=0$ ($\omega_{de}=-1$) and
$\dot H<0$ ($\omega_{de}>-1$). More specifically, at the transition
$\omega_{de}=\omega_{de}^{eff}=-1$, we require
\begin{equation}
b=-\frac{3(1+A)}{n(1+r_x)}\left( 1-\frac{X}{\rho_{de}^{1+\alpha}}
\right).
\end{equation}
Since $0\leq b\leq1$ and $n>-3$ for $\dot{\Lambda}>0$, we obtain a
restriction $-3<n<0$ from Eq. (34). Lastly we take
$\Lambda(t)=C_6\rho_{de}^\epsilon$ hence we get
\begin{equation}
\omega_{de}^{eff}=-1+\frac{3(1+A)}{3+n}\left(
1-\frac{X}{\rho_{de}^{1+\alpha}}
\right)-\frac{C_6\epsilon\rho_{de}^{\epsilon-1}\dot\rho_{de}}{6(3+n)H^3\Omega_{de}}+\frac{nb(1+r_x)}{n+3}.
\end{equation}
which we can be simplified to yield
\begin{equation}
\omega_{de}^{eff}=-1+\frac{1+A}{3+n}\left(3+\frac{\epsilon
C_6\rho_{de}^\epsilon}{2H^2\Omega_{de}}\right)\left(
1-\frac{X}{\rho_{de}^{1+\alpha}} \right)+\frac{nb(1+r_x)}{n+3}.
\end{equation}
At the epoch of phantom crossing, we require
$\omega_{de}=-1=\omega_{de}^{eff}$ to get
\begin{equation}
b=-\frac{1+A}{n(1+r_x)}\left(3+\frac{\epsilon
C_6\rho_{de}^\epsilon}{2H^2\Omega_{de}}\right)\left(
1-\frac{X}{\rho_{de}^{1+\alpha}} \right)
\end{equation}
Note that Eq. (37) is reduced to (34) if $\epsilon=0$.

\newpage
\section{Conclusion and discussion}

In the present work we attempted to resolve the cosmic coincidence
problem in the context of five-dimensional brane world gravitation
theory. The CCP is fairly alleviated since stable stationary
solution exists for the dynamical system. The dark energy is
represented by the modified Chaplygin gas and is further assumed to
interact with the matter. This interaction leads to a phantom
crossing scenario. We have also determined various effective EoS for
dark energy using different choices of $\Lambda(t)$, since EoS of
dark energy will change if it interacts with matter. This paper also
presents a generalization of the work in \cite{wu} where it is shown
that the dark energy with the MCG EoS crosses the phantom divide in
the background of four dimensional FRW spacetime.

Models of interacting dark energy have taken considerable interest
in recent years. The model mimics $\Lambda$CDM at early times while
it gives a finite dark energy-dark matter ratio at late times so
that coincidence problem is alleviated. It turns out that dark
energy domination is merely a transient event and will be replaced
by the dark matter dominant era once again \cite{cabral}. It is
recently shown that this model can resolve the cosmic age problem as
well since simple dark energy cannot remove the problem
\cite{wang1}. The cosmic age predicted by the interacting model is
predicted to be greater than the $\Lambda$CDM model which is
consistent with the observations and hence alleviates the cosmic age
problem. Further, this model is also investigated in the context of
loop quantum cosmology, a theory in which all cosmological
singularities are avoided due to quantum effects \cite{chen}. It is
also of some interest that the coupling parameter for the
interaction yields a variable Newton's gravitational constant
\cite{zen}. From the observational point of view, a small but
non-vanishing interaction is reported from the analysis of the
dynamics of 33 relaxed galaxy clusters like Abell A586
\cite{bertolami}.

\subsubsection*{Acknowledgment}
I am grateful to M.R. Setare, D. Polarski, G.S. Khadekar and V.K.
Onemli for useful comments on this work.

\end{document}